\documentclass[11pt,oneside,reqno]{amsart}
\usepackage[latin9]{inputenc}
\usepackage{mathrsfs}
\usepackage{amsthm}
\usepackage{amsbsy}
\usepackage{amstext}
\usepackage{amssymb}
\usepackage{esint}

\makeatletter
\numberwithin{equation}{section}
\numberwithin{figure}{section}
\theoremstyle{plain}
\newtheorem{thm}{\protect\theoremname}[section]
  \theoremstyle{remark}
  \newtheorem{rem}[thm]{\protect\remarkname}

\@ifundefined{date}{}{\date{}}

\def\QED{\hskip0.1em\hfill\null\ \null\nobreak\hfill
\kern3pt\lower1.8pt\vbox{\hrule\hbox
{\vrule\kern1pt\vbox{\kern1.7pt \hbox{$\scriptstyle
QED$}\kern0.2pt}\kern1pt\vrule}\hrule}}

\numberwithin{equation}{section}

\renewcommand{\mathcal}[1]{\mathscr{#1}}

\newcommand{\ie}{\textit{i.e.}}
\newcommand{\eg}{\textit{e.g.}}

\makeatother

  \providecommand{\remarkname}{Remark}
\providecommand{\theoremname}{Theorem}

\begin{document}
\global\long\def\ga{\alpha}
\global\long\def\gb{\beta}
\global\long\def\ggm{\gamma}
\global\long\def\go{\omega}
\global\long\def\ge{\epsilon}
\global\long\def\gs{\sigma}
\global\long\def\gd{\delta}
\global\long\def\gD{\Delta}
\global\long\def\vph{\varphi}
\global\long\def\gf{\varphi}
\global\long\def\gk{\kappa}
\global\long\def\wh#1{\widehat{#1}}
\global\long\def\bv#1{\mathbf{#1}}
\global\long\def\bs#1{\boldsymbol{#1}}
\global\long\def\ui{\wh{\boldsymbol{\imath}}}
\global\long\def\uj{\wh{\boldsymbol{\jmath}}}
\global\long\def\uk{\wh{\textbf{\em k}}}
\global\long\def\bosy#1{\boldsymbol{#1}}
\global\long\def\vect#1{\overline{\mathbf{#1}}}
\global\long\def\uI{\widehat{\mathbf{I}}}
\global\long\def\uJ{\widehat{\mathbf{J}}}
\global\long\def\uK{\widehat{\mathbf{K}}}
\global\long\def\uv#1{\widehat{\mathbf{#1}}}
\global\long\def\cross{\boldsymbol{\times}}
\global\long\def\ddt{\frac{\dee}{\dee t}}
\global\long\def\dbyd#1{\frac{\dee}{\dee#1}}
\global\long\def\parby#1#2{\frac{\partial#1}{\partial#2}}
\global\long\def\fall{,\quad\text{for all}\quad}
\global\long\def\reals{\mathbb{R}}
\global\long\def\rthree{\reals^{3}}
\global\long\def\rsix{\reals^{6}}
\global\long\def\les{\leqslant}
\global\long\def\ges{\geqslant}
\global\long\def\dee{\mathrm{\mathrm{d}}}
\global\long\def\from{\colon}
\global\long\def\tto{\longrightarrow}
\global\long\def\abs#1{\left|#1\right|}
\global\long\def\isom{\cong}
\global\long\def\comp{\circ}
\global\long\def\cl#1{\overline{#1}}
\global\long\def\fun{\varphi}
\global\long\def\interior{\mathrm{Int\,}}
\global\long\def\diver{\mathrm{div\,}}
\global\long\def\sign{\mathrm{sign\,}}
\global\long\def\dimension{\mathrm{dim\,}}
\global\long\def\esssup{\mathrm{ess}\,\sup}
\global\long\def\ess{\mathrm{{ess}}}
\global\long\def\kernel{\text{Kernel}\,}
\global\long\def\support{\mathrm{Supp}\,}
\global\long\def\image{\mathrm{Image\,}}
\global\long\def\resto#1{|_{#1}}
\global\long\def\incl{\iota}
\global\long\def\rest{\rho}
\global\long\def\extnd{e_{0}}
\global\long\def\proj{\pi}
\global\long\def\sphere{S^{2}}
\global\long\def\hemis{H}
\global\long\def\ino#1{\int\limits _{#1}}
\global\long\def\half{\frac{1}{2}}
\global\long\def\shalf{{\scriptstyle \half}}
\global\long\def\third{\frac{1}{3}}
\global\long\def\empt{\varnothing}
\global\long\def\paren#1{\left(#1\right)}
\global\long\def\bigp#1{\bigl(#1\bigr)}
\global\long\def\biggp#1{\biggl(#1\biggr)}
\global\long\def\Bigp#1{\Bigl(#1\Bigr)}
\global\long\def\braces#1{\left\{  #1\right\}  }
\global\long\def\sqbr#1{\left[#1\right]}
\global\long\def\norm#1{\|#1\|}
\global\long\def\trps{^{\mathsf{T}}}
\global\long\def\alt{\mathfrak{A}}
\global\long\def\pou{\eta}
\global\long\def\ext{\bigwedge}
\global\long\def\forms{\Omega}
\global\long\def\dotwedge{\dot{\mbox{\ensuremath{\wedge}}}}
\global\long\def\vel{\theta}
\global\long\def\contr{\raisebox{0.4pt}{\mbox{\ensuremath{\lrcorner}}}\,}
\global\long\def\fcontr{\raisebox{0.4pt}{\mbox{\ensuremath{\llcorner}}}\,}
\global\long\def\lie{\mathcal{L}}
\global\long\def\jet#1{j^{1}(#1)}
\global\long\def\Jet#1{J^{1}(#1)}
\global\long\def\L#1{L\bigl(#1\bigr)}
\global\long\def\vvforms{\ext^{\dims}\bigp{T\spc,\vbts^{*}}}
\global\long\def\contr{\raisebox{0.4pt}{\mbox{\ensuremath{\lrcorner}}}\,}
\global\long\def\lisub#1#2#3{#1_{1}#2\dots#2#1_{#3}}
\global\long\def\lisup#1#2#3{#1^{1}#2\dots#2#1^{#3}}
\global\long\def\lisubb#1#2#3#4{#1_{#2}#3\dots#3#1_{#4}}
\global\long\def\lisubbc#1#2#3#4{#1_{#2}#3\cdots#3#1_{#4}}
\global\long\def\lisubbwout#1#2#3#4#5{#1_{#2}#3\dots#3\widehat{#1}_{#5}#3\dots#3#1_{#4}}
\global\long\def\lisubc#1#2#3{#1_{1}#2\cdots#2#1_{#3}}
\global\long\def\lisupc#1#2#3{#1^{1}#2\cdots#2#1^{#3}}
\global\long\def\lisupp#1#2#3#4{#1^{#2}#3\dots#3#1^{#4}}
\global\long\def\lisuppc#1#2#3#4{#1^{#2}#3\cdots#3#1^{#4}}
\global\long\def\lisuppwout#1#2#3#4#5#6{#1^{#2}#3#4#3\wh{#1^{#6}}#3#4#3#1^{#5}}
\global\long\def\lisubbwout#1#2#3#4#5#6{#1_{#2}#3#4#3\wh{#1}_{#6}#3#4#3#1_{#5}}
\global\long\def\lisubwout#1#2#3#4{#1_{1}#2\dots#2\widehat{#1}_{#4}#2\dots#2#1_{#3}}
\global\long\def\lisupwout#1#2#3#4{#1^{1}#2\dots#2\widehat{#1^{#4}}#2\dots#2#1^{#3}}
\global\long\def\lisubwoutc#1#2#3#4{#1_{1}#2\cdots#2\widehat{#1}_{#4}#2\cdots#2#1_{#3}}
\global\long\def\twp#1#2#3{\dee#1^{#2}\wedge\dee#1^{#3}}
\global\long\def\thp#1#2#3#4{\dee#1^{#2}\wedge\dee#1^{#3}\wedge\dee#1^{#4}}
\global\long\def\fop#1#2#3#4#5{\dee#1^{#2}\wedge\dee#1^{#3}\wedge\dee#1^{#4}\wedge\dee#1^{#5}}
\global\long\def\idots#1{#1\dots#1}
\global\long\def\icdots#1{#1\cdots#1}
\global\long\def\pis{x}
\global\long\def\pib{X}
\global\long\def\body{B}
\global\long\def\man{\mathscr{M}}
\global\long\def\bdry{\partial}
\global\long\def\gO{\varOmega}
\global\long\def\reg{R}
\global\long\def\bdom{\bdry\reg}
\global\long\def\bndo{\partial\gO}
\global\long\def\pbndo{\Gamma}
\global\long\def\bndoo{\pbndo_{0}}
 \global\long\def\bndot{\pbndo_{t}}
\global\long\def\cloo{\cl{\gO}}
\global\long\def\nor{\boldsymbol{n}}
\global\long\def\nora{\nor}
\global\long\def\norb{\boldsymbol{u}}
\global\long\def\norc{v}
\global\long\def\dA{\,\dee A}
\global\long\def\dV{\,\dee V}
\global\long\def\eps{\varepsilon}
\global\long\def\vs{\mathbf{W}}
\global\long\def\avs{\mathbf{V}}
\global\long\def\vbase{\boldsymbol{e}}
\global\long\def\sbase{\mathbf{e}}
\global\long\def\vf{w}
\global\long\def\avf{u}
\global\long\def\stn{\varepsilon}
\global\long\def\rig{r}
\global\long\def\rigs{\mathcal{R}}
\global\long\def\qrigs{\!/\!\rigs}
\global\long\def\qd{\!/\,\!\kernel\diffop}
\global\long\def\dis{\chi}
\global\long\def\fc{F}
\global\long\def\st{\sigma}
\global\long\def\bfc{b}
\global\long\def\sfc{t}
\global\long\def\stm{S}
\global\long\def\sts{\varSigma}
\global\long\def\ebdfc{T}
\global\long\def\optimum{\st^{\mathrm{opt}}}
\global\long\def\scf{K}
\global\long\def\curr{T}
\global\long\def\forms{\Omega}
\global\long\def\cee#1{C^{#1}}
\global\long\def\lone{L^{1}}
\global\long\def\linf{L^{\infty}}
\global\long\def\lp#1{L^{#1}}
\global\long\def\ofbdo{(\bndo)}
\global\long\def\ofclo{(\cloo)}
\global\long\def\vono{(\gO,\rthree)}
\global\long\def\vonbdo{(\bndo,\rthree)}
\global\long\def\vonbdoo{(\bndoo,\rthree)}
\global\long\def\vonbdot{(\bndot,\rthree)}
\global\long\def\vonclo{(\cl{\gO},\rthree)}
\global\long\def\strono{(\gO,\reals^{6})}
\global\long\def\sob{W_{1}^{1}}
\global\long\def\sobb{\sob(\gO,\rthree)}
\global\long\def\lob{\lone(\gO,\rthree)}
\global\long\def\lib{\linf(\gO,\reals^{12})}
\global\long\def\ofO{(\gO)}
\global\long\def\oneo{{1,\gO}}
\global\long\def\onebdo{{1,\bndo}}
\global\long\def\info{{\infty,\gO}}
\global\long\def\infclo{{\infty,\cloo}}
\global\long\def\infbdo{{\infty,\bndo}}
\global\long\def\ld{LD}
\global\long\def\ldo{\ld\ofO}
\global\long\def\ldoo{\ldo_{0}}
\global\long\def\trace{\gamma}
\global\long\def\pr{\proj_{\rigs}}
\global\long\def\pq{\proj}
\global\long\def\qr{\,/\,\reals}
\global\long\def\aro{S_{1}}
\global\long\def\art{S_{2}}
\global\long\def\mo{m_{1}}
\global\long\def\mt{m_{2}}
\global\long\def\yieldc{B}
\global\long\def\yieldf{Y}
\global\long\def\trpr{\pi_{P}}
\global\long\def\devpr{\pi_{\devsp}}
\global\long\def\prsp{P}
\global\long\def\devsp{D}
\global\long\def\ynorm#1{\|#1\|_{\yieldf}}
\global\long\def\colls{\Psi}
\global\long\def\ssx{S}
\global\long\def\smap{s}
\global\long\def\smat{\chi}
\global\long\def\sx{e}
\global\long\def\snode{P}
\global\long\def\elem{e}
\global\long\def\nel{L}
\global\long\def\el{l}
\global\long\def\ipln{\phi}
\global\long\def\ndof{D}
\global\long\def\dof{d}
\global\long\def\nldof{N}
\global\long\def\ldof{n}
\global\long\def\lvf{\chi}
\global\long\def\lfc{\varphi}
\global\long\def\amat{A}
\global\long\def\snomat{E}
\global\long\def\femat{E}
\global\long\def\tmat{T}
\global\long\def\fvec{f}
\global\long\def\snsp{\mathcal{S}}
\global\long\def\slnsp{\Phi}
\global\long\def\ro{r_{1}}
\global\long\def\rtwo{r_{2}}
\global\long\def\rth{r_{3}}
\global\long\def\mind{\alpha}
\global\long\def\vb{\xi}
\global\long\def\vbt{E}
\global\long\def\fib{\mathbf{V}}
\global\long\def\jetb#1{J^{#1}}
\global\long\def\jetm#1{j_{#1}}
\global\long\def\sobp#1#2{W_{#2}^{#1}}
\global\long\def\inner#1#2{\left\langle #1,#2\right\rangle }
\global\long\def\fields{\sobp pk(\vb)}
\global\long\def\bodyfields{\sobp p{k_{\partial}}(\vb)}
\global\long\def\forces{\sobp pk(\vb)^{*}}
\global\long\def\bfields{\sobp p{k_{\partial}}(\vb\resto{\bndo})}
\global\long\def\loadp{(\sfc,\bfc)}
\global\long\def\strains{\lp p(\jetb k(\vb))}
\global\long\def\stresses{\lp{p'}(\jetb k(\vb)^{*})}
\global\long\def\diffop{D}
\global\long\def\strainm{E}
\global\long\def\incomps{\vs_{\yieldf}}
\global\long\def\devs{L^{p'}(\eta_{1}^{*})}
\global\long\def\incompsns{L^{p}(\eta_{1})}
\global\long\def\dists{\mathcal{D}'}
\global\long\def\testfs{\mathcal{D}}
\global\long\def\prop{P}
\global\long\def\aprop{Q}
\global\long\def\flux{T}
\global\long\def\fform{\tau}
\global\long\def\dimn{n}
\global\long\def\sdim{{\dimn-1}}
\global\long\def\prodf{{\scriptstyle \smallsmile}}
\global\long\def\ptnl{\varphi}
\global\long\def\form{\omega}
\global\long\def\dens{\rho}
\global\long\def\simp{s}
\global\long\def\cell{C}
\global\long\def\chain{B}
\global\long\def\ach{A}
\global\long\def\coch{X}
\global\long\def\scale{s}
\global\long\def\fnorm#1{\abs{#1}^{\flat}}
\global\long\def\chains{\mathcal{A}}
\global\long\def\ivs{\boldsymbol{U}}
\global\long\def\mvs{\boldsymbol{V}}
\global\long\def\cvs{\boldsymbol{W}}
\global\long\def\subbs{\mathcal{B}}
\global\long\def\elements{\mathcal{E}}
\global\long\def\element{E}
\global\long\def\nodes{\mathcal{N}}
\global\long\def\node{N}
\global\long\def\psubbs{\mathcal{P}}
\global\long\def\psubb{P}
\global\long\def\matr{M}
\global\long\def\nodemap{\nu}
\global\long\def\prop{p}
\global\long\def\radi{K}
\global\long\def\radvec{\mathbf{i}}
\global\long\def\radint{i}
\global\long\def\mrad{\mathbf{I}}
\global\long\def\mradint{I}
\global\long\def\irrad{\phi}
\global\long\def\Hemis{\mathscr{H}}
\global\long\def\phasesp{\rthree\times\sphere}
\global\long\def\sterad{\omega}
\global\long\def\meters{w}
\global\long\def\coneu{\hat{u}}
\global\long\def\moment{M}
\global\long\def\pdists{U}
\global\long\def\prerad{\kappa}
\global\long\def\pmd{v}
\global\long\def\tradi{J}
\global\long\def\measrs{M}
\global\long\def\total{\Phi}
\global\long\def\mradi{J}
\global\long\def\R{\reals}
\global\long\def\Z{\mathbb{Z}}

\title{Geometric Aspects of Singular Dislocations}

\author{Marcelo Epstein and Reuven Segev \\
 }

\address{University of Calgary, Canada; Ben-Gurion University, Israel}

\subjclass[2000]{74A05; 74E20; 58A25}

\keywords{Continuum mechanics; dislocations; differential forms; singularities;
de Rham currents; Frank's rules.}
\begin{abstract}
The theory of singular dislocations is placed within the framework
of the theory of continuous dislocations using de Rham currents. For
a general $n$-dimensional manifold, an $(n-1)$-current describes
a local layering structure and its boundary in the sense of currents
represents the structure of the dislocations. Frank's rules for dislocations
follow naturally from the nilpotency of the boundary operator.

\bigskip{}

\end{abstract}
\maketitle

\section{Introduction}

The aim of this work is to establish a precise relationship between
the theory of continuous distributions of defects and its discrete
counterpart. Historically, the latter was developed first by pioneers
like Volterra and Somigliana. The continuous theory was arrived at
later by, among others, Bilby \cite{Bilby55},  Kr\"oner \cite{kroener},
Kondo \cite{kondo} and Noll \cite{noll}. Methodologically speaking,
the passage from the discrete to the continuous theory was perhaps
spurred by the realization that certain differential geometric objects
already provide a heuristic path to generalize the discrete, intuitively
graspable, picture. The clearest example is provided by the lack of
closure of a Burgers' circuit enclosing an edge dislocation in two
dimensions. The picture of this event so much resembles that of the
lack of commutativity of two vector fields, that one would be remiss
to ignore the analogy. And, in fact, the analogy is in this case fully
justifiable on physical grounds. In the infinitesimal limit, the lack
of closure alluded to above is a lack of integrability, whose intensity
is measured by the torsion of a distant parallelism associated with
the smeared-out underlying crystal lattice. Once this particular instance
was exploited, the temptation could not be resisted to attribute some
putative physical meaning to all kinds of other geometric objects,
from Riemann curvatures to Einstein tensors. The works of Kondo and,
later, Noll inaugurated the emergence of the opposite paradigm. Instead
of building the theory, as it were, from the bottom up, they adopted
the puristic tenet that the presence of defects (or inhomogeneities)
in a continuum should be encoded automatically within the constitutive
equations of the body and only there, without any spurious intervention
of an atomic substrate. The two approaches do not necessarily lead
to the same results.

Since both approaches mentioned above have undeniable merits, it
would be futile to argue for one to the exclusion of the other. Our
intention is, therefore, to place the theory of singular dislocations,
\ie, dislocations that are concentrated rather than distributed continuously,
within the elegant framework of the smooth theory and so to come full
circle to the original physical picture. The mathematical tool for
this purpose consists of a weak reinterpretation of the differential
geometric objects of the smooth theory in terms of the geometric theory
of de Rham currents \cite{derham,federer}.

The framework presented below applies to general manifolds. In particular,
no Riemannian structure is assumed. The theory of integration of differential
forms on manifolds is used throughout and the necessary background
may be found in \cite{epsgeom,segev} within the context of continuum
mechanics and in standard references on differential geometry, such
as those cited therein. Specifically, for the smooth case, we consider
a local layering of the material as represented geometrically by a
$1$-form $\vph$, the layering form, on a manifold $\man$ which
may represent a material body or its image under a configuration in
space. The condition that the body contains no dislocations is thus
represented mathematically by the local integrability condition $\dee\vph=0$.
This condition implies that at least locally (and also globally if
$\man$ is contractible to a point) $\vph$ is the gradient of a function
$u$. The layers, \ie,  hypersurfaces of constant values of $u$,
may be thought of as deformed Bravais planes of a crystalline body.
Dislocations are present in regions where $\dee\vph\ne0$ so that
such a system of layers is not available.

In the non-smooth situation, we generalize the layering $1$-form
to a de Rham $(n-1)$-current $\curr$ and the condition for the body
to have no dislocations is generalized to $\bdry\curr=0$, \ie,
that the boundary of the current vanishes. It is noted that in his
exposition on singularities in the deformations of solids, Cermelli
\cite{Cermelli99} makes use of de Rham currents.

It is interesting to note that the language of differential forms
and currents rather than that of frame fields, enables one to analyze
dislocations associated with a single layering form or current (rather
than three). From the physical point of view this means that a single
family of Bravais planes is sufficient for the study of the possible
presence of dislocations. It is also noteworthy that Frank's conservation
rule%
\footnote{In the words of \cite[p. 34]{read}, Frank's rules are: 1. ``The Burgers
vector is conserved along a dislocation''; 2. ``The sum of the Burgers
vectors of the dislocations meeting at a node must vanish.''%
} follows naturally from a basic mathematical property of currents,
specifically, the vanishing of the boundary of a boundary.

\section{The smooth theory of crystal dislocations}

\label{sec:parallelism}

\subsection{Frames and coframes}

The simplest theory of continuous distributions of dislocations assumes
that a material body $\mathcal{B}$ is endowed with a distinguished
smooth field of bases of its tangent spaces. In general, for an $n$-dimensional
manifold $\mathcal{M}$, such a field can be regarded as a section
$\sigma$ of the frame bundle $F{\mathcal{M}}$, namely
\begin{eqnarray}
\sigma:{\mathcal{M}} & \tto & F{\mathcal{M}}\nonumber \\
x & \longmapsto & \sigma(x)=\{{\bf e}_{1},...,{\bf e}_{n}\},
\end{eqnarray}
with $\pi\circ\sigma=\textrm{id}_{\mathcal{M}}$, where $\pi$ is
the bundle projection and $\textrm{id}_{\mathcal{M}}$ is the identity
map in $\mathcal{M}$. We remark that smooth, or even continuous,
global sections may not exist in general. A manifold for which a smooth
section of $F{\mathcal{M}}$ does exist is said to be \textit{parallelizable},
and the field of basis induces a \textit{teleparallelism} (or \textit{distant
parallelism}) in the manifold.

There is an alternative (dual) way to look at a distant parallelism.
Indeed, any \textit{frame} (that is, any basis of the tangent space
$T_{x}{\mathcal{M}}$) induces a unique dual basis (or \textit{coframe})
of the cotangent space $T_{x}^{*}{\mathcal{M}}$, and vice versa.
Therefore, a distant parallelism can also be seen as the choice of
a particular coframe field. Put differently, a distant parallelism
induces an ${\R}^{n}$-valued one-form on ${\mathcal{M}}$. Moreover,
this coframe field can be regarded as a cross section $\sigma^{*}$
of the coframe bundle of $\mathcal{M}$.

Let $x^{i}\;(i=1,...,n)$ be a coordinate system. Then, the frame
field is given by:
\begin{equation}
{\bf e}_{\alpha}={e}_{\;\alpha}^{i}\frac{\partial}{\partial x^{i}},\qquad\alpha=1,...,n,\label{weak2}
\end{equation}
where the matrix with entries ${e}_{\;\alpha}^{i}={e}_{\;\alpha}^{i}(x^{1},...x^{n})$
is nowhere singular. Accordingly, the corresponding coframe is given
by
\begin{equation}
{\bf e}^{\alpha}={e}_{\; i}^{\alpha}\dee x^{i},\;\;\;\;\alpha=1,...,n,\label{weak3}
\end{equation}
where ${e}_{\; i}^{\alpha}$ are the entries of the inverse matrix.
As already pointed out, a coframe is clearly represented as a collection
of $n$ pointwise linearly independent $1$-forms indexed by $\alpha$.

\subsection{The interpretation of covectors as Bravais hyperplanes}

\label{sec:bravais}

Covectors may be used to describe geometrically collections of hyperplanes.
Let ${\bf W}$ be a finite dimensional vector space and let $f\in{\bf W}^{*}$
be a covector, that is, a linear functional $f:{\bf W}\to{\R}$. Then,
$f$ may be identified uniquely with the collection $H_{f1}$ of vectors
$w$ in ${\bf W}$ such that $f(w)=1$. This collection of vectors
constitutes a hyperplane. The difference between any two elements
of $H_{f1}$ belongs to the kernel $H_{f0}$ of the operator $f$.
Clearly, parallel hyperplanes $H_{fk}$ will be obtained if we consider
the elements $w\in{\bf W}$ such that $f(w)=k$ for any integer $k$.
Conversely, for any $w\in{\bf W}$, $f(w)$ may be interpreted as
the amount of hyperplanes that the arrow representing $w$ penetrates.
If a covector is multiplied by a positive number $a$, the density
of the planes is multiplied by $a$. A covector includes a choice
of orientation for the various planes (a positive side versus a negative
side of a plane) and the multiplication of $f$ by $-1$ reverses
the orientation.

It is natural therefore, to use a covector as a continuous model of
a system of parallel planes in a crystal (Bravais) lattice. In fact,
if $\vs=\rthree$, for a covector $f$ given as
\begin{equation}
f=f_{j}\dee x^{i},
\end{equation}
relative to the standard dual basis $\dee x^{i}$, the components
$(f_{1},f_{2},f_{3})$ are proportional to the Miller indices for
the system of parallel planes. Whereas the Miller indices are normalized
to provide a direction only, the covector $f$ contains additional
information as to the density of the layers. Let $n$ be the dimension
of ${\bf W}$. Then, a collection of $n$ linearly independent covectors
$\{f^{\alpha}\},\;\alpha=1,\dots,n$, induces a collection of $n$
families of parallel hyperplanes. These covectors, therefore, represent
a Bravais lattice.

So far we considered a single vector space ${\bf W}$. On a manifold
$\mathcal{M}$, the interpretation just described can be applied in
a point-wise manner, that is, to each tangent space $T_{x}{\mathcal{M}}$.
Thus, at each point $x\in{\mathcal{M}}$, the covector representing
the ``direction'' and density of the layers is given by the value
$\phi(x)$ of a 1-form $\phi:{\mathcal{M}}\to T^{*}{\mathcal{M}}$.
Noting that this setting does not require any additional structure,
metric or otherwise, it is natural to study the geometric properties
of such a differential form as a representation of the structure of
a class of layers in a lattice. We will refer to such a form as a
\emph{local layering form}.

\subsection{Integrability}

\subsubsection{Intuitive considerations \label{sub:Intuitive-considerations}}

Let ${\bf F}(x)=F^{i}(x){\bf e}_{i}$ be a vector field in a 3-dimensional
Euclidean vector space with Cartesian coordinates $(x^{1},x^{2},x^{3})$,
where $\{{\bf e}_{i}\}$ is an orthonormal basis. We recall that the
condition that the vector field be conservative, namely, that there
exists a scalar potential function $u(x)$ such that
\begin{equation}
F^{i}=\frac{\partial u}{\partial x^{i}},\label{intu1}
\end{equation}
is
\begin{equation}
\frac{\partial F^{i}}{\partial x^{j}}-\frac{\partial F^{j}}{\partial x^{i}}=0,\label{intu2}
\end{equation}
for all $i,j=1,2,3$. The above well-known scheme may be generalized
to an arbitrary differentiable manifold $\mathcal{M}$ using the terminology
of differential forms. In particular, one says that a differentiable
$r$-form $\phi$ is \textit{exact} if there is an $(r-1)$-form $\alpha$
such that $\phi=\dee\alpha$, where $\dee$ denotes the \textit{exterior
differentiation} of forms. Thus, $\alpha$ is a potential form for
$\phi$. Every exact form $\phi$ is automatically \textit{closed},
that is $\dee\phi=0$, since the $\dee$ operator enjoys the property
$\dee^{2}=0$. Conversely, if the manifold $\mathcal{M}$ is \textit{contractible
to a point}, every closed form $\phi$ is exact, that is, derives
from a potential. In the general case, when the manifold is not necessarily
contractible to a point, if $\phi$ is closed, for each $x\in{\mathcal{M}}$
there is a neighborhood where $\phi$ is exact. Thus, the condition
that the form be closed is a generalization of the condition (\ref{intu2})
for the existence of a local potential.

In particular, a 1-form $\phi$ on an $n$-dimensional manifold $\man$
has $n$ components, just as a vector field. Let $\phi$ be represented
locally as
\begin{equation}
\phi=\phi_{i}\dee x^{i},\label{intu3}
\end{equation}
where now $x^{i}$ is a manifold coordinate patch. Its exterior derivative
is the 2-form $\dee\phi$ represented locally by
\begin{equation}
\begin{split}\dee\phi=\sum\limits _{i,j}\phi_{i,j}\dee x^{j}\wedge\dee x^{i} & =\half\left(\sum\limits _{i,j}\phi_{i,j}\dee x^{j}\wedge\dee x^{i}+\sum\limits _{i,j}\phi_{j,i}\dee x^{i}\wedge\dee x^{j}\right),\\
 & =\half\left(\sum\limits _{i,j}\phi_{i,j}\dee x^{j}\wedge\dee x^{i}-\sum\limits _{i,j}\phi_{j,i}\dee x^{j}\wedge\dee x^{i}\right),\\
 & =\half\sum\limits _{i,j}\left(\phi_{i,j}-\phi_{j,i}\right)\dee x^{j}\wedge\dee x^{i}.,
\end{split}
\end{equation}
where commas indicate partial derivatives. Thus, the condition $\dee\phi=0$
is the analog
\begin{equation}
\phi_{i,j}-\phi_{j,i}=0\label{intu5}
\end{equation}
of Equation (\ref{intu2}).

We conclude that a closed 1-form represents a locally coherent collection
of layers that may be identified with deformed lattice planes. In
general, the $2$-form $\delta=\dee\vph$ is a measure of the nature
of dislocation density and we will refer to it as the \emph{dislocation
density form}.

Let $Z$ be a $2$-dimensional manifold of $\man$ with boundary $Y=\bdry Z$.
Then, by Stokes' theorem
\begin{equation}
I=\int_{Y}\vph=\int_{Z}\gd\label{eq:ConserveTotalDislocation}
\end{equation}
It follows that $\int_{Y}\vph$ is independent of the particular submanifold
$Z$. If there exists a submanifold $Z_{0}$ on which $\dee\vph=0$,
\ie, there are no dislocations on $Z_{0}$, then, $\int_{Z}\gd=0$
for any other submanifold $Z$ with boundary $Y$, even if $Z$ passes
through a region where dislocations exist (\ie, $\gd\ne0$). In the
general case where no such manifold $Z_{0}$ exists, $\int_{Z}\gd$
is still independent of $Z$, and $I$ above is a measure the total
dislocation embraced by $Y$ in analogy with the Burgers vector.

\subsubsection{Parallelism and coordinate systems}

In the definition of our distant parallelism, the differentiability
of the cross section has played no role whatsoever. If, on the other
hand, $\sigma^{*}$ is differentiable, we may calculate its exterior
derivative $\dee\sigma^{*}$. In components, we obtain
\begin{equation}
\tau^{\alpha}=\dee{\bf e}^{\alpha}={e}_{\;\; i,j}^{\alpha}\;\dee x^{j}\wedge\dee x^{i},\;\;\;\;\alpha=1,...,n,\label{weak4}
\end{equation}
namely, $\tau=\dee\sigma^{*}$ is an ${\R}^{n}$-valued two-form which
we call the \textit{torsion of the parallelism}.%
\footnote{Notice an inessential difference with the usual definition, whereby
the form takes values in $T_{x}{\mathcal{M}}$ rather than in ${\R}^{3}$.%
} The identical vanishing of the torsion form, namely,
\begin{equation}
\tau_{ij}^{\alpha}=0\;\;\;\;\;\alpha,i,j=1,...,n,\label{weak4a}
\end{equation}
is necessary and sufficient for the existence of a local coordinate
system such that the original frame field $\sigma$ becomes its natural
base. From the point of view of the theory of dislocations in continuous
media, if the frames represent crystalline bases, the vanishing of
the torsion implies that the body can be smoothly brought to a configuration
in which all crystal bases within a coordinate patch are mutually
parallel, so that there are no defects in the lattice. The body is
then \textit{locally homogeneous.} Conversely, a non-vanishing torsion
is an indication, and perhaps a measure, of the dislocation density
(or \textit{inhomogeneity}). For comprehensive treatments of the general
theory of inhomogeneity see \eg,  \cite{wang,epselz}.

In the Bravais-lattice interpretation of Section \ref{sec:bravais},
one may ask whether, given a 1-form $\phi$, there is a ``potential''
function $u:{\mathcal{M}}\to\R$ such that $\phi=\dee u$, where $\dee$
denotes the exterior derivative (which is identical to the gradient
in this situation). Such a potential function, if it exists, will
label the various layers, at least locally. Indeed, these layers would
be precisely the (local) level surfaces of this potential. If each
one of a collection of $n$ 1-forms $\{\phi^{\alpha}\}$, whose values
at each point $x\in{\mathcal{M}}$ are linearly independent covectors,
derives from a local potential function $u^{\alpha}$, the values
$u^{\alpha}(x_{0})$ represent a point $x_{0}\in{\mathcal{M}}$ uniquely
and the body has acquired locally the crystalline structure of a perfect
(non-dislocated) Bravais lattice. Recalling that the condition for
the existence of a local potential $u^{\alpha}$ for a 1-form $\phi^{\alpha}$
is the equality of the cross-derivatives, \ie,
\begin{equation}
\frac{\partial\phi_{i}^{\alpha}}{\partial x^{j}}=\frac{\partial\phi_{j}^{\alpha}}{\partial x^{i}},
\end{equation}
we recover the integrability condition (\ref{weak4a}).

As just indicated, the vanishing of the torsion forms $\tau_{ij}^{\alpha}$
is necessary for the integrability (holonomicity, homogeneity) of
the coframe field ${\bf e}^{\alpha}$. In terms of the original frame
field ${\bf e}_{\alpha}$, on the other hand, it is well known that
the existence of an adapted coordinate system is guaranteed by the
commutativity of each pair of base vector fields, namely,
\begin{equation}
L_{{\bf e}_{\alpha}}{\bf e}_{\beta}=[{\bf e}_{\alpha},{\bf e}_{\beta}]=0,\label{weak4b}
\end{equation}
where $L_{{\bf u}}{\bf v}$ is the \textit{Lie derivative} of the
vector field ${\bf v}$ in the direction of the vector field ${\bf u}$
and where $[{\bf u},{\bf v}]$ denotes their \textit{Lie bracket}.
In terms of components, this can be written as
\begin{equation}
\frac{\partial{{e}_{\beta}^{i}}}{\partial x^{j}}\;{e}_{\alpha}^{j}-\frac{\partial{{e}_{\alpha}^{i}}}{\partial x^{j}}\;{e}_{\beta}^{j}=0.\label{weak4c}
\end{equation}
Since
\begin{equation}
0=\frac{\partial({e}_{\beta}^{i}\;{e}_{j}^{\beta})}{\partial x^{k}}=\frac{\partial{e}_{\beta}^{i}}{\partial x^{k}}\;{e}_{j}^{\beta}+\frac{{e}_{j}^{\beta}}{\partial x^{k}}\;{e}_{\beta}^{i},\label{weak4d}
\end{equation}
the Lie bracket can be expressed as
\begin{equation}
[{\bf e}_{\alpha},{\bf e}_{\beta}]=\left(\tau_{ij}^{\sigma}{e}_{\alpha}^{i}{e}_{\beta}^{j}\right){\bf e}_{\sigma}.\label{weak4e}
\end{equation}

In the context of dislocations, we may identify the Lie bracket between
the two base vector fields ${\bf e}_{\alpha}$ and ${\bf e}_{\beta}$
as the local \textit{Burgers vector} ${\bf B}_{\alpha\beta}$ between
the corresponding crystal directions. Expressing the Burgers vector
${\bf B}_{\alpha\beta}$ in the local basis, we may distinguish between
the \textit{edge component} of the dislocation density and its \textit{screw
component}, the former being the part contained in the plane spanned
by the base vectors ${\bf e}_{\alpha}$ and ${\bf e}_{\beta}$.

\subsection{Frank's rule as the vanishing of the boundary of a boundary}

Starting from the notions of simplices and chains in an affine space
and their generalization to manifolds, one arrives at a fundamental
geometrical and topological result of clear intuitive meaning. It
states that the boundary $\partial U$ of any well-defined domain
of integration $U$ must necessarily have a vanishing boundary, namely,
\begin{equation}
\partial^{2}U=\partial\partial U=0.\label{frank1}
\end{equation}

In the theory of smooth differential forms, on the other hand, the
operation of exterior differentiation enjoys a formally similar property,
namely, for any differential form $\omega$ on a manifold
\begin{equation}
\dee^{2}\omega=\dee\dee\omega=0.\label{frank2}
\end{equation}

The relation and consistency between these two identities is mediated
by \textit{Stokes' theorem},
\begin{equation}
\int\limits _{U}\dee\omega=\int\limits _{\partial U}\omega,\label{frank3}
\end{equation}
where $\omega$ is an arbitrary $(p-1)$-form and $U$ is an arbitrary
$p$-dimensional domain of integration.

We will presently show that a smooth version of Frank's rule for dislocation
branching, \cite{frank} (see also \cite{read}), can be obtained
as a direct consequence of these purely geometric identities. For
dimension 3, we observe that, since for any given smooth coframe field
${\bf e}^{\alpha}$ the torsion $\tau^{\alpha}=d{\bf e}^{\alpha}$
consists of 3 exact 2-forms, the integral of the torsion over the
boundary of any $3$-dimensional domain of integration $U$ vanishes,
\begin{equation}
\int\limits _{\partial U}\tau^{\alpha}=0.\label{weak4f}
\end{equation}
In the physical interpretation, this implies that there are no isolated
dislocation sources, not even as a smoothed-out approximation. In
particular, consider a tubular domain $U$ and an arbitrarily small
neighborhood $V$ of $U$ such that the torsion vanishes in $V\smallsetminus U$,
then, if we intercept $U$ transversely by means of two (oppositely
oriented) lids $\Sigma_{1}$ and $\Sigma_{2}$ giving rise to a finite
tube $\hat{U}$, we obtain
\begin{equation}
0=\int\limits _{\hat{U}}\dee\tau^{\alpha}=\int\limits _{\partial{\hat{U}}}\tau^{\alpha}=\int\limits _{\Sigma_{1}}\tau^{\alpha}-\int\limits _{\Sigma_{2}}\tau^{\alpha}.\label{weak4g}
\end{equation}
In other words, the integral of the torsion over any tube cross section
is constant. The same reasoning can be applied to a tube with branches,
thus providing the smooth version of Frank's rule. We note that the
restriction of the coframe field to $V\smallsetminus U$ consists
of three closed 1-forms, by construction. But, since any curve surrounding
the tube is not contractible to a point, these 1-forms are not necessarily
exact. In physical terms, the body minus the tube is only locally
homogeneous. Moreover, the integral along any such non-contractible
curve of the coframe 1-forms gives rise to three constants, each of
them exactly equal to the integral of the corresponding $\alpha$-component
of the torsion over any cross section.

From the heuristic point of view, as the diameter of the tube shrinks,
we may impose the condition that the torsion increases proportionately
so as to keep its integral over the cross section constant and thus
recover the classical form of Frank's rule. The rigorous mathematical
treatment of this limiting process will be handled in the sequel using
the language of currents.

\section{The weak counterpart}

So as to generalize the notions just introduced, we define a \textit{weak
teleparallelism} $\rho$ as an ${\R}^{n}$-valued \textit{current}
in the sense of de Rham \cite{derham,federer}. We recall that a de
Rham $r$-current is a linear functional on the vector space of $C^{\infty}$
differential $r$-forms with compact supports in $\man$ such that
$\curr(\phi)\to0$ if the components of $\phi$ and all their derivatives
tend to zero uniformly in the support of $\phi$. A de Rham current
is the natural generalization of a Schwartz distribution to manifolds
where the forms $\phi$ are analogous to test functions. As such,
currents provide a tool for the study of non-smooth, concentrated,
physical phenomena, \eg, dislocation lines and slip surfaces. We
identify naturally an $n$-tuple of $r$-currents with an $\reals^{n}$-valued
$r$-current.

\subsection{The current induced by a form\label{sub:CurrentByAForm}}

A $1$-form $\vph$ on $\man$ may be paired with a smooth $(n-1)$-form
$\psi$, having a compact support, to produce a real number in the
form
\begin{equation}
\int_{\man}\vph\wedge\psi.\label{eq:PairingForms}
\end{equation}
Here, $\vph\wedge\psi$ denotes the exterior product of the two forms,
an $n$-form having a compact support which may be integrated over
the $n$-dimensional manifold $\man$. Thus, the form $\vph$ induces
a linear functional $\curr_{\vph}$ acting on the vector space of
$(n-1)$-forms of compact supports in $\man$ in the form
\begin{equation}
\curr_{\vph}(\psi)=\int_{\man}\vph\wedge\psi.\label{eq:currActionFromAForm}
\end{equation}
If all the derivatives of the local representatives of the form $\psi$
tend uniformly to zero in compact subsets of the domains of charts
in $\man$, then $\curr_{\vph}(\psi)$ tends to zero. It follows that
$\curr_{\vph}$ is indeed an $(n-1)$-current.
\begin{rem}
The action (\ref{eq:currActionFromAForm}) may be given a physical
interpretation in a different context. The 1-form $\vph$ may be interpreted
as a force field per unit value of a certain extensive property. For
example, as the electric field in the case where the property under
consideration is the electric charge. Thus, the question whether $\vph$
is closed corresponds to the question of the existence of a potential
function for the force field. The $(n-1)$-form $\psi$ is interpreted
as the flux field of the property under consideration so that for
any $n$-dimensional region $\reg\subset\man$,
\begin{equation}
\int_{\bdry\reg}\psi
\end{equation}
is interpreted as the total flux of the property through the boundary
$\bdom$. Thus, $\curr_{\vph}(\psi)$ in (\ref{eq:currActionFromAForm})
may be interpreted as the total power expended by the force field
while the transport of the property is given by the flux field $\psi$.
\end{rem}

\begin{rem}
\label{InterpretationAsStresses}A collection of $n$ $1$-forms in
$\reals^{n}$, \eg, $\{{\bf e}^{1},...,{\bf e}^{n}\}$ may be interpreted
as the collection of gradients of the components of a velocity field.
The corresponding velocity field will be incompatible, or contain
dislocation rates, if the forms are not closed. Thus, for a collection
of $n$ $(n-1)$-forms $\psi^{i}$ the action
\begin{equation}
\int_{\man}\sum_{i}\sbase^{i}\wedge\psi^{i}
\end{equation}
may be interpreted as the mechanical power performed by the stress
matrix having the components $\psi^{i}$ (each $\psi^{i}$ has $n$
components itself) on the velocity gradient.
\end{rem}

\subsection{Examples of currents in general\label{sub:Examples-of-Currents}}

An $(n-1)$-current in the form (\ref{eq:currActionFromAForm}) is
very special as it is induced by a smooth $1$-form. As such it may
be identified with the form $\vph$. The fine topology used for the
test forms enables one to define currents which are a lot less regular.
Such currents, generalizing the local layering forms, will be referred
to as \emph{local layering currents.} We present below a number of
examples.

\subsubsection{\label{sub:CurrentByForm}The current induced by a form}

We have seen already that a current $\curr_{\vph}$, induced by a
closed $1$-form $\vph$ according to Equation (\ref{eq:currActionFromAForm}),
represents a locally coherent system of layers as in Section \ref{sub:Intuitive-considerations}
.

\subsubsection{\label{sub:Incoherence}Incoherence at an interface}

Let $\man=\reals^{2}=\{(x^{1},x^{2})\}$ and consider the $1$-form
\begin{equation}
\vph(x)=\begin{cases}
\dee x^{1}, & x^{2}<0,\\
a\dee x^{1}, & a>0\in\reals,\, x^{2}\ges0.
\end{cases}\label{eq:StepCurr}
\end{equation}
The form $\vph$ is not continuous, yet the current $\curr_{\vph}$
as in Equation (\ref{eq:currActionFromAForm}) is well defined. For
$x^{2}<0$, $\vph$ represents a collection of vertical layers and
for $x^{2}>0$, $\vph$ represents a collection of vertical layers
that are $a$ times more dense. Thus, the form $\vph$ describes incoherence
at the interface $x^{2}=0$ for $a\ne1$.

\subsubsection{A Dirac current}

Let $\lisub v,{n-1}$ be a collection of vectors in $T_{x_{0}}\man$
for a point $x_{0}\in\man$. Then,
\begin{equation}
T(\psi)=\psi(x_{0})(\lisub v,{n-1})
\end{equation}
is an $(n-1)$-current which is a generalization of the Dirac delta
distribution.

\subsubsection{\label{sub:CurrentByDPhi}The current induced by the exterior derivative}

Consider the $(n-2)$-current $T_{\dee\vph}$ induced by a $1$-form
$\vph$ as
\begin{equation}
T_{\dee\vph}(\form)=\int_{\man}\dee\vph\wedge\form,\label{eq:boundarySmoothForms}
\end{equation}
for any compactly supported $(n-2)$-form $\form$.  Using the basic
property of exterior differentiation whereby
\begin{equation}
\dee(\ga\wedge\gb)=\dee\ga\wedge\gb+(-1)^{r}\ga\wedge\dee\gb,
\end{equation}
 for an $r$-form $\ga$, Equation (\ref{eq:boundarySmoothForms})
may be written in the form
\begin{equation}
\begin{split}T_{\dee\vph}(\form) & =\int_{\man}\dee(\vph\wedge\form)+\int_{\man}\vph\wedge\dee\form,\\
 & =\int_{\bdry\man}\vph\wedge\form+\int_{\man}\vph\wedge\dee\form,\\
 & =\int_{\man}\vph\wedge\dee\form,
\end{split}
\label{eq:BoundarySmoothForms-a}
\end{equation}
where in the second line we used Stokes's theorem and in the third
line we used the fact that $\form$ is compactly supported in $\man$.

\subsubsection{\label{sub:Polyhedral-chains}Polyhedral chains as currents }

A current $\curr_{\simp}$ can be uniquely associated with an $(n-1)$-simplex
$\simp$ in $\man$. It is defined as
\begin{equation}
\curr_{\simp}(\psi)=\int_{\simp}\psi,
\end{equation}
for any compactly supported $(n-1)$-form $\psi$. Thus, rather than
a continuous system of layers modeled by a form $\vph$ as in Section
\ref{sub:CurrentByForm} above, $\curr_{\simp}$ represents a single
``concentrated'' layer. For example, a simplex $\simp$ inside $\man$
may represent a cut inside the body where an additional layer of atoms
has been added or removed.\par Evidently, we may extend this definition
to an arbitrary chain $A=\sum_{p}a_{p}\simp_{p}$ and define $\curr_{A}$
by
\begin{equation}
\curr_{A}(\psi)=\sum_{p}a_{p}\int_{\simp_{p}}\psi.
\end{equation}

\subsubsection{\label{sub:ProductFunctionCurrent}The product of a current by a
function}

Let $\curr$ be a current and $u$ a smooth function. Then, one may
define the product current $u\curr$ by
\begin{equation}
u\curr(\psi)=\curr(u\psi).
\end{equation}
In particular, for the current $\curr_{\simp}$ of the previous example,
\begin{equation}
u\curr(\psi)=\int_{\simp}u\psi.
\end{equation}

\subsection{Dislocations as boundaries of currents\label{sub:Boundaries-of-Currents}}

We recall that the boundary of a $p$-current $\curr$ is the $(p-1)$-current
$\bdry\curr$ defined by
\begin{equation}
\bdry\curr(\form)=\curr(\dee\form).
\end{equation}
In case $\bdry\curr=0$, one says that $\curr$ is closed. Just as
a current is a non-smooth generalization of the system of layers represented
by a $1$-form, $\bdry\curr=0$ is a generalization of the condition
$\dee\vph=0$ (see Section \ref{sub:BoundaryTphi} below) implying
coherence of the system. In fact, a theorem by de Rham (see \cite[pp. 79--80]{derham})
asserts that any closed current is homologous to a current $\curr_{\vph}$
induced by some smooth form $\vph$, \ie, there exists a current
$S$ such that for each compactly supported smooth form $\psi$,
\begin{equation}
(\curr-\curr_{\vph})(\psi)=\curr(\psi)-\int_{\man}\vph\wedge\psi=\bdry S(\psi).
\end{equation}
In other words, the homological properties of $\curr$ may be obtained
by the analogous properties of a current induced by an approximating
smooth form $\vph$. As the dislocation structure of a smooth form
$\vph$ is obtained by $\dee\vph$ and as $\bdry\curr_{\vph}=\curr_{\dee\vph}$
(Section \ref{sub:BoundaryTphi}), it is natural to obtain the dislocation
structure induced by the $(n-1)$-current $\curr$ from its boundary
$\bdry\curr$. Thus we will refer to the $(n-2)$-current $D=\bdry\curr$
as the \emph{dislocation current}.

Again, we demonstrate the significance of these notions in the following
examples.

\subsubsection{\label{sub:BoundaryTphi}The boundary of a current induced by a smooth
form}

Consider Section \ref{sub:CurrentByDPhi} above. It follows from Equation
(\ref{eq:BoundarySmoothForms-a}) that for a $1$-form $\vph$,
\begin{equation}
\bdry\curr_{\vph}=\curr_{\dee\vph}.
\end{equation}
We conclude that if $\vph$ is closed then $\bdry\curr_{\vph}=0$.
This is just the condition for coherence phrased in terms of currents.

\subsubsection{The dislocation current for an incoherent interface}

Consider Example 2 above. Using $\reals^{2-}$ and $\reals^{2+}$
to denote the lower and upper half planes in $\reals^{2}$ , we have
for each $0$-form $\form$,
\begin{equation}
\begin{split}\bdry\curr_{\vph}(\form) & =\int_{\reals^{2-}}\dee x^{1}\wedge\dee\form+\int_{\reals^{2+}}a\dee x^{1}\wedge\dee\form,\\
 & =-\int_{\reals^{2-}}\dee(\dee x^{1}\wedge\form)+\int_{\reals^{2-}}\dee^{2}x^{2}\wedge\form\\
 & \qquad-\int_{\reals^{2+}}a\dee(\dee x^{1}\wedge\form)+\int_{\reals^{2+}}a\dee^{2}x^{2}\wedge\form,\\
 & =-\int_{\bdry\reals^{2-}}\dee x^{1}\wedge\form-\int_{\bdry\reals^{2+}}a\dee x^{1}\wedge\form,\\
 & =(a-1)\int_{\bdry\reals^{2-}}\form dx^{1},
\end{split}
\end{equation}
where it is noted that $\bdry\reals^{2-}$ and $\bdry\reals^{2+}$
contain the set $L=\{(x^{1},0)\}$ but with opposite orientations.
Let $\curr_{L}$ be the $0$-current in $\reals^{2}$ defined by
\begin{equation}
\curr_{L}(\form)=\int_{\bdry\reals^{2-}}\form dx^{1}.
\end{equation}
Then, the preceding calculation shows that
\begin{equation}
\bdry\curr_{\vph}=(a-1)\curr_{L}.
\end{equation}
Indeed, for the case where $a=1$, $\bdry\curr_{\vph}=0$ and $\curr_{\vph}$
is a closed current which represents a coherent collection of layers.
In case $a\ne1$, the dislocations are concentrated on the line $L$
which is the support of $\bdry\curr_{\vph}$, \ie, $\bdry\curr_{\vph}(\form)=0$
for any $0$-form (a function) $\form$ whose support is disjoint
from $L$.

\subsubsection{The dislocation current induced by a polyhedral chain}

Referring to Section \ref{sub:Polyhedral-chains}, we note that by
Stokes's theorem,
\begin{equation}
\begin{split}\bdry\curr_{\simp}(\form) & =\int_{\simp}\dee\form,\\
 & =\int_{\bdry\simp}\form,
\end{split}
\end{equation}
so that
\begin{equation}
\bdry\curr_{\simp}=\curr_{\bdry\simp},
\end{equation}
where $\bdry\simp$ is viewed as a polyhedral chain. Thus, as one
would expect, the dislocation line is the boundary of the embedded
simplex. Evidently, the boundary operator is linear and may be extended
in this case to a polyhedral chain.

\subsubsection{General incoherent interfaces}

Let $Y$ be an $n$-dimensional submanifold of $\man$ with boundary
$Z=\bdry Y$. Let $\vph$ be a closed form and consider the layering
current
\begin{equation}
\curr(\psi)=\int_{Y}a\vph\wedge\psi+\int_{\bar{Y}}\vph\wedge\psi,
\end{equation}
where $a>0\in\reals$ and $\bar{Y}$ is the manifold with boundary
$-Z$ whose interior is $\man\smallsetminus Y$ (so that the orientation
of $\bdry(\man\smallsetminus Y)$ is the opposite of the orientation
of $Z$). We have
\begin{equation}
\begin{split}D(\form)=\bdry\curr(\form) & =\int_{Y}a\vph\wedge\dee\form+\int_{\bar{Y}}\vph\wedge\dee\form,\\
 & =-\int_{Y}a\dee(\vph\wedge\form)+\int_{Y}a\dee\vph\wedge\form,\\
 & \quad-\int_{\bar{Y}}\dee(\vph\wedge\form)+\int_{\bar{Y}}\dee\vph\wedge\form,
\end{split}
\end{equation}
and using the assumption that $\dee\vph=0$, it follows that
\begin{equation}
\bdry\curr(\form)=(1-a)\int_{Z}\vph\wedge\form.
\end{equation}
Thus, the dislocations are distributed over the boundary of $Y$ while
the material is coherent inside and outside $Y$.

\subsubsection{A dislocation line}

Consider the case where $\man=(-1,1)^{3}$ is an open cube in $\rthree$.
Let
\begin{equation}
\simp=\{(0,x^{2},x^{3})\in\man\mid\, x^{2}\les0\}
\end{equation}
equipped with the orientation induced by the form $\dee x^{2}\wedge\dee x^{3}$
and let $\curr_{\simp}$ be the $2$-current defined by
\begin{equation}
\curr_{\simp}(\vph)=\int_{\simp}\vph
\end{equation}
for any $2$-form $\vph$ compactly supported in $\man$. It follows
that for any compactly supported $1$-form $\form$,
\begin{equation}
\begin{split}\bdry\curr_{\simp}(\form) & =\int_{\simp}\dee\form,\\
 & =\int_{L}\form,
\end{split}
\end{equation}
where $L=\{(0,0,x^{3})\in\man\}$ oriented naturally by the form $\dee x^{3}$.
Clearly, $L$ represents the line of dislocation associated with the
half plane $\simp$. Notice how closely the layering current $\curr_{\simp}$
matches the addition of a half plane of atoms as depicted in standard
texts on dislocations.

\subsubsection{The boundary of a product of a function and a chain}

Using again the setting of Section \ref{sub:ProductFunctionCurrent},
let $s$ be an $(n-1)$-simplex in the $n$-dimensional manifold $\man$
and let $u$ be a smooth function. Set
\begin{equation}
\curr_{u\simp}(\psi)=\int_{\simp}u\psi
\end{equation}
for every compactly supported $(n-1)$-form $\psi$ on $\man$. It
follows that for every compactly supported $(n-2)$-form $\form$
on $\man$ one has
\begin{equation}
\begin{split}\bdry\curr_{u\simp}(\form) & =\curr_{u\simp}(\dee\form),\\
 & =\int_{\simp}u\dee\form,\\
 & =\int_{\simp}\dee(u\form)-\int_{\simp}\dee u\wedge\form,\\
 & =\int_{\bdry\simp}u\form-\int_{\simp}\dee u\wedge\form,\\
 & =(\curr_{u\bdry\simp}-T_{\simp}\fcontr\dee u)(\form).
\end{split}
\end{equation}
Here, we have used the notation
\begin{equation}
\curr\fcontr\ga(\form)=\curr(\ga\wedge\form)
\end{equation}
for an $r$-current $\curr$, a $p$-form $\ga$ and a compactly supported
smooth $(r-p)$-form $\form$. Thus, in general, one has
\begin{equation}
\bdry\curr_{u\simp}=\curr_{u\bdry\simp}-T_{\simp}\fcontr\dee u.\label{eq:bdryOfProduct}
\end{equation}
Clearly, one may replace $\simp$ above by a polyhedral chain or a
smooth submanifold (using triangulation).

\subsubsection{A node of three dislocation lines}

Let $\man=(-1,1)^{3}\subset\rthree$, $\simp_{1}=\{(0,x^{2},x^{3})\in\man\mid x^{2}\les0,\, x^{3}\ges0\}$,
$\simp_{2}=\{(x^{1},x^{2},0)\in\man\mid x^{1}\ges0,\, x^{2}\les0\}$,
$\simp_{3}=\{(x^{1},x^{2},0)\in\man\mid x^{1}\les0,\, x^{2}\les0\}$
where the quarter planes $\simp_{1}$, $\simp_{2}$, $\simp_{3}$
are oriented by the normals $\nor_{1}=(1,0,0)$, $\nor_{2}=(0,0,1)$
and $\nor_{3}=(0,0,-1)$, respectively. Consider the layering described
by the current
\begin{equation}
\curr=\curr_{a_{1}\simp_{1}}+\curr_{a_{2}\simp_{2}}+\curr_{a_{3}\simp_{3}},
\end{equation}
so that
\begin{equation}
\curr(\psi)=\sum_{i=1}^{3}a_{i}\int_{\simp_{i}}\psi,
\end{equation}
for a smooth compactly supported $2$-form $\psi$. Thus,
\begin{equation}
\begin{split}D(\form) & =\bdry\curr(\form),\\
 & =\sum_{i=1}^{3}a_{i}\int_{\simp_{i}}\dee\form,\\
 & =\sum_{i=1}^{3}a_{1}\int_{\bdry\simp_{i}}\form,\\
 & =(a_{1}\curr_{L_{1}}+a_{2}\curr_{L_{2}}+a_{3}\curr_{L_{3}}+(a_{1}-a_{2}-a_{3})\curr_{L})(\form),
\end{split}
\end{equation}
where the one dimensional simplices $L_{p}$ are define as follows:
$L_{1}$ is the segment from the origin to $(0,0,1)$, $L_{2}$ is
the segment from $(1,0,0)$ to the origin, $L_{3}$ is the segment
from $(-1,0,0)$ to the origin, $L$ is the segment form $(0,-1,0)$
to the origin. It is noted immediately that if the dislocation current
is supported only on the ``fork'' $L_{1}\cup L_{2}\cup L_{3}$,
then, one has the condition
\begin{equation}
a_{1}=a_{2}+a_{3}.
\end{equation}
Evidently, this result, a particular case of Frank's second rule,
will also hold if the cube is deformed under any embedding in $\rthree$.
Furthermore, the choice of planes is immaterial. (See also Section
\ref{sub:Frank's-second-rule}.)

\subsection{The boundary of a boundary and Frank's rules}

The theory of currents provides a generalization of the intuitive
result of combinatorial topology that the boundary of the boundary
of a chain is zero. This follows immediately from
\begin{equation}
\bdry^{2}\curr(\ga)=\bdry\curr(\dee\ga)=\curr(\dee^{2}\ga)=0.\label{eq:bdryOfBdry}
\end{equation}

The dislocation current $D$ is obtained as the boundary of a current
$T$. Hence,
\begin{equation}
\bdry D=\bdry^{2}\curr=0
\end{equation}
is a condition that the dislocation current must satisfy. In other
words, the dislocation current must be closed.

We may use this result in the following situations.

\subsubsection{Frank's first rule}

Let $L$ be an $(n-2)$-dimensional submanifold without boundary in
the manifold $\man$. (For example, in a three dimensional situation,
$L$ could be a curve that does not have ends inside $\man$.) We
assume that $L$ is the support of the dislocation current $D$. We
want to examine the possibility that the dislocation current is of
the form
\begin{equation}
D=\curr_{uL}
\end{equation}
for some real valued function $u$ defined on $\man$. Thus, there
is a local layering $(n-1)$-current $S$ such that
\begin{equation}
D=\curr_{uL}=\bdry S.
\end{equation}
Using Equation (\ref{eq:bdryOfProduct}) for the submanifold $L$
(instead of $\simp$), we have
\begin{equation}
\begin{split}0 & =\bdry D,\\
 & =\bdry\curr_{uL},\\
 & =\curr_{u\bdry L}-\curr_{L}\fcontr\dee u,
\end{split}
\end{equation}
and, by the assumption that $\bdry L=0$, we conclude that
\begin{equation}
\dee u=0
\end{equation}
so that $u$ must be constant on $L$. This result is clearly analogous
to the Frank's first rule (in which case $L$ is $1$-dimensional)
and it is an example of the constancy theorem of geometric measure
theory.

\subsubsection{Frank's first rule for the boundary of a submanifold}

Let $Z$ be an $(n-1)$-dimensional submanifold with boundary $\bdry Z$
of $\man$, let $u$ be a smooth function on $\man$ and consider
the $(n-1)$-current $\curr_{uZ}$ given as
\begin{equation}
\curr_{uZ}(\psi)=\int_{Z}u\psi.
\end{equation}
Using the analog of (\ref{eq:bdryOfProduct}) for the submanifold
$Z$, we have
\begin{equation}
\bdry\curr_{uZ}=\curr_{u\bdry Z}-T_{Z}\fcontr\dee u.
\end{equation}
Assume that $\bdry\curr_{uZ}$ is supported on $\bdry Z$. By applying
$\bdry T_{uZ}$ to forms whose supports are disjoint from $\bdry Z$,
it follows that $\dee u$ must vanish on $Z$. We conclude therefore
that if $u$ is not constant on $Z$, the support $\bdry T_{uZ}$
contains points outside $\bdry Z$. In the context of dislocations,
if the ``intensity'' of the dislocations along a certain line $L$
is not constant, there should be additional continuous dislocations
on the dislocation surface $Z$ outside $L$.

\subsubsection{\label{sub:Frank's-second-rule}Frank's second rule}

Let $\man$ be a nonempty bounded open subset of $\rthree$ and let
$A\in\man$. Consider $3$ curves $L_{i}$, $i=1,2,3$ in $\man$
such that each $L_{i}$ is connected in $\man$, it originates at
$A$ and is the intersection of the image of a curve $c_{i}:[0,1]\to\rthree$
with $\man$ such that $c_{i}(1)\notin\man$. (In other words, each
$c_{i}$ ends on the topological boundary of $\man$ where it is noted
that $\man$ has no boundary as a manifold and not as a current.)
Thus, $\bdry L_{i}=\{A\}$ as a manifold and $\bdry\curr_{L_{i}}=T_{A}$
(the Dirac delta) as a current. We examine the case where the dislocation
current is given by $D=\sum_{i}a_{i}T_{L_{i}}$. It follows that for
an arbitrary compactly supported smooth $0$-form $\ga$,
\begin{equation}
\begin{split}0 & =\bdry D(\ga),\\
 & =\sum_{i}a_{i}\int_{L_{i}}\dee\ga,\\
 & =\sum_{i}a_{i}\ga(A).
\end{split}
\end{equation}
We conclude therefore that $0=\sum a_{i}$. This condition is evidently
analogous to Frank's second rule for dislocations.


\begin{thebibliography}{Wan67}

\bibitem[BBS55]{Bilby55}
B.A. Bilby, R.~Bullough, and E.~Smith.
\newblock Continuous distributions of dislocations: A new application of the
  methods of non-{R}iemannian geometry.
\newblock {\em Proceedings of the Royal Society of London}, A 231:263--273,
  1955.

\bibitem[Cer99]{Cermelli99}
P.~Cermelli.
\newblock Material symmetry and singularities in solids.
\newblock {\em Proceedings of the Royal Society of London}, A 455:299--322,
  1999.

\bibitem[dR84]{derham}
G.~de~Rham.
\newblock {\em Differentiable Manifolds}.
\newblock Springer, 1984.

\bibitem[EE07]{epselz}
M.~Elzanowski and M.~Epstein.
\newblock {\em Material Inhomogeneities and their Evolution}.
\newblock Springer, 2007.

\bibitem[Eps10]{epsgeom}
M.~Epstein.
\newblock {\em The Geometrical Language of Continuum Mechanics}.
\newblock Cambridge University Press, 2010.

\bibitem[Fed69]{federer}
H.~Federer.
\newblock {\em Geometric Measure Theory}.
\newblock Springer, 1969.

\bibitem[Fra51]{frank}
F.C. Frank.
\newblock Crystal dislocations--{E}lementary concepts and definitions.
\newblock {\em Philosophical Magazine}, 42:809--819, 1951.

\bibitem[Kon55]{kondo}
K.~Kondo.
\newblock {\em Geometry of Elastic Deformation and incompatibility}.
\newblock Tokyo Gakujutsu Benken Fukyu-Kai, IC, 1955.

\bibitem[Kr{\"o}59]{kroener}
E.~Kr{\"o}ner.
\newblock Allgemeine {K}ontinuumstheorie der {V}ersetzungen und
  {E}igenspannungen.
\newblock {\em Archive for Rational Mechanics and Analysis}, 4:273--334, 1959.

\bibitem[Nol67]{noll}
W.~Noll.
\newblock Materially uniform bodies with inhomogeneities.
\newblock {\em Archive for Rational mechanics and Analysis}, 27:1--32, 1967.

\bibitem[Rea53]{read}
W.T. Read.
\newblock {\em Dislocations in Crystals}.
\newblock McGraw-Hill, 1953.

\bibitem[Seg12]{segev}
R.~Segev.
\newblock Notes on metric independent analysis of classical fields.
\newblock {\em Mathematical Methods in the Applied Sciences}, 2012.
\newblock DOI: 10.1002/mma.2610.

\bibitem[Wan67]{wang}
C.-C. Wang.
\newblock On the geometric structure of simple bodies, a mathematical
  foundation for the theory of continuous distributions of dislocations.
\newblock {\em Archive for Rational Mechanics and Analysis}, 27:33--94, 1967.

\end{thebibliography}

\end{document}